\documentclass[10pt,journal]{IEEEtran}
\usepackage{graphicx}
\usepackage{color}

\begin{document}
\title{An Intraoperative $\beta^-$ Detecting Probe
For Radio-Guided Surgery in Tumour Resection}

\author{
\IEEEauthorblockN{
Andrea~Russomando\IEEEauthorrefmark{1}\IEEEauthorrefmark{2}\IEEEauthorrefmark{3},
Fabio~Bellini\IEEEauthorrefmark{1}\IEEEauthorrefmark{2}, 
Valerio~Bocci\IEEEauthorrefmark{2},
Giacomo~Chiodi\IEEEauthorrefmark{2}, 
Francesco~Collamati\IEEEauthorrefmark{2}\IEEEauthorrefmark{4}, 
Erika~De~Lucia\IEEEauthorrefmark{5},
Raffaella~Donnarumma\IEEEauthorrefmark{1},
Riccardo~Faccini\IEEEauthorrefmark{1}\IEEEauthorrefmark{2},
Carlo~Mancini-Terracciano\IEEEauthorrefmark{1}\IEEEauthorrefmark{2},
Michela~Marafini\IEEEauthorrefmark{2}\IEEEauthorrefmark{6},
Riccardo~Paramatti\IEEEauthorrefmark{2}, 
Vincenzo~Patera\IEEEauthorrefmark{2}\IEEEauthorrefmark{4}\IEEEauthorrefmark{6},
%Davide~Pinci\IEEEauthorrefmark{2},  
Luigi~Recchia\IEEEauthorrefmark{2}, 
Alessio~Sarti\IEEEauthorrefmark{4}\IEEEauthorrefmark{5},
Adalberto~Sciubba\IEEEauthorrefmark{2}\IEEEauthorrefmark{4}\IEEEauthorrefmark{6},
Elena~Solfaroli~Camillocci\IEEEauthorrefmark{1}\IEEEauthorrefmark{2},
Cecilia~Voena\IEEEauthorrefmark{2} and
Silvio~Morganti\IEEEauthorrefmark{2}
}
%affiliations
\thanks{
\IEEEauthorblockA{\IEEEauthorrefmark{1}
Dip. Fisica, Sapienza Univ. di Roma, Rome, Italy;}
\IEEEauthorblockA{\IEEEauthorrefmark{2}
INFN Sezione di Roma, Rome, Italy;}
\IEEEauthorblockA{\IEEEauthorrefmark{3}
Center for Life Nano Science@Sapienza, Istituto Italiano di Tecnologia, Rome, Italy.}
\IEEEauthorblockA{\IEEEauthorrefmark{4}
Dip. Scienze di Base e Applicate per l'Ingegneria, Sapienza Univ. di Roma, Rome, Italy;}
\IEEEauthorblockA{\IEEEauthorrefmark{5}
Laboratori Nazionali di Frascati dell’INFN, Frascati, Italy;}
\IEEEauthorblockA{\IEEEauthorrefmark{6}
Museo Storico della Fisica e Centro Studi e Ricerche ``E. Fermi'', Rome, Italy;}
}}
\maketitle
\pagestyle{empty}
\thispagestyle{empty}

\begin{abstract}
The development of the $\beta^-$ based radio-guided surgery
aims to extend the technique to 
those tumours where surgery is the only possible treatment 
and the assessment of the resection would most profit 
from the low background around the lesion, as for brain tumours.
Feasibility studies on meningioma, glioma, and neuroendocrine tumors
already estimated the potentiality of this new treatment.
To validate the technique, prototypes of the intraoperative probe required by the technique to detect $\beta^-$ radiation have been developed. This paper discusses the design details of the device and the tests performed in laboratory.
In such tests particular care has to be taken to reproduce the surgical field conditions.
The innovative technique to produce  specific phantoms and the dedicated testing protocols is described in detail.\end{abstract}

%\begin{IEEEkeywords}

%\end{IEEEkeywords}

\section{Introduction}
\IEEEPARstart{T}{he} main advantage of 
the radio-guided surgery (RGS) technique 
exploiting $\beta^-$ radiation~\cite{Beta-RGS,PatentBeta-RGS},
compared to the traditional RGS 
using $\gamma$ radiation~\cite{GammaRGS},
is a more favourable  ratio between the signal coming from the tumor and the rest of the body.
Conversely to the tracers with $\gamma$ emitters,
pure $\beta^-$ radionuclides emit electrons
which penetrate only a few millimeters of tissue 
and produce almost no gamma radiation (the 
\textit{bremsstrahlung} contribution above 100~keV being less than 0.1\%)
resulting in a very low background on the lesion signal.

Operating in a low background environment %, instead,
allows the development of a handy and compact probe which,
detecting particles emitted locally,
provides a clearer delineation of margins of the lesioned tissue. 
The RGS technique applicability can therefore be extended
also to cases with large uptake of the tracer
from healthy organs next to the lesion, 
minimizing the radiotracer activity to be administered.
Low exposure of the medical team is also expected~\cite{Beta-RGS}.

For this technique to enter in the clinical practice, relevant clinical cases need to be identified, suitable radio-tracers need to be  developed for them and a detector needs to be designed  and tested for this specific application.

The most relevant clinical cases are those where the RGS with $\gamma$ radiation is not applicable because of the presence of uptaking organs nearby, for instance cerebral, abdominal and pediatric tumors. 
\begin{figure}[!tbh]
\centering
 \includegraphics [width=0.48\textwidth]{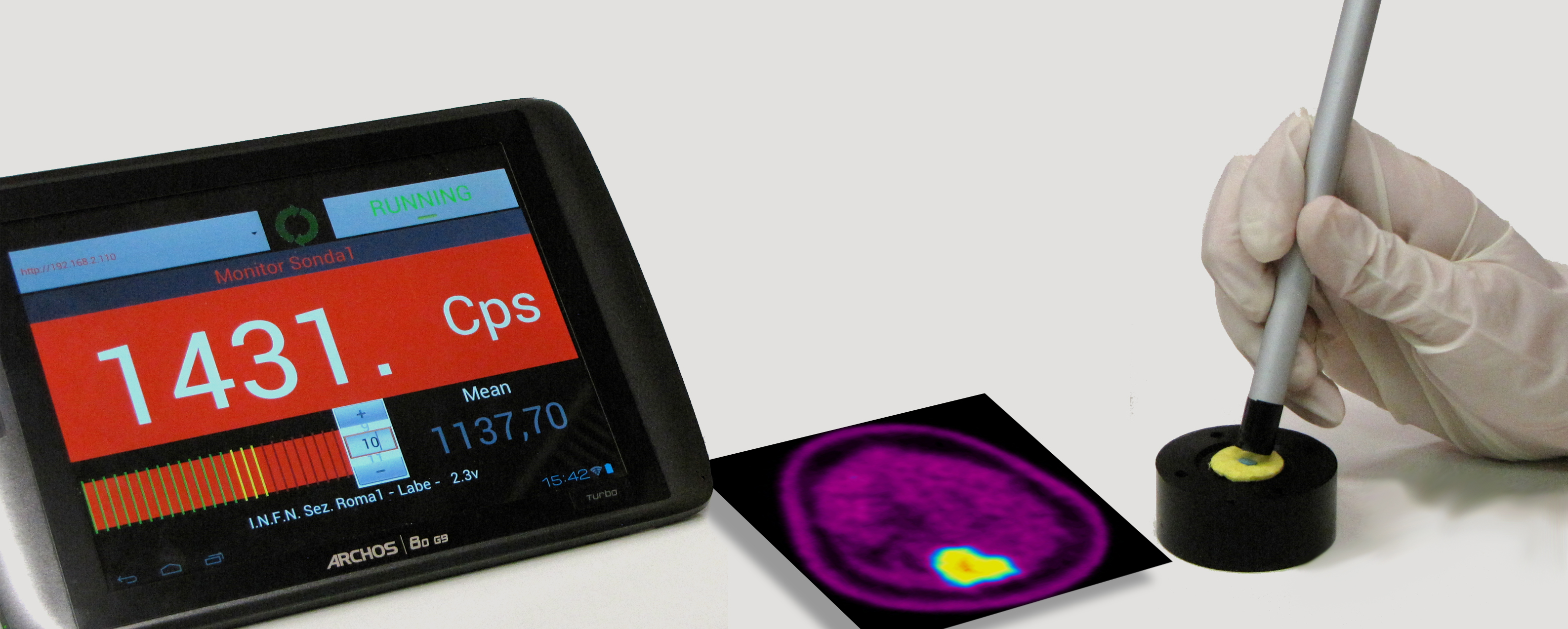}
 \caption{Image of the intraoperative $\beta^-$ probe.}
 \label{fig:probeS4}
 \end{figure}

Among the existing radio-tracers those that are linked to pure $\beta^-$ emitters are those used in radio-methabolic treatments with $^{90}$Y, such as Peptide Receptor Radionuclide Therapy~\cite{PRRT}. Such tracers are somatostatin analogues (e.g. DOTATOC~\cite{90Y-DOTATOC}) and  very selective for meningioma~\cite{MeningiomaUptake} and  neuroendocrine tumors (NET)~\cite{NETUptake}. Our first choice has been indeed to use meningioma as proof of principle  and abdominal NETs as a first application with clinical relevance.  In the case of administration of $^{90}$Y-DOTATOC for meningioma, glioma and NET we have performed feasibility studies~\cite{MeningiomaRGS,RENET} of which the present paper describes critical ingredients. We are finalizing the authorizations for the clinical tests. Nonetheless, to extend this technique beyond NETs, other radio-tracers need to be  considered and we cannot restrict only to pure $\beta^-$ decays. There is indeed a significant number of nuclides which have a half-life ($\tau_{1/2}$) compatible with clinical applications and a limited $\gamma$ radiation. For instance $^{186}$Re has $\tau_{1/2}\sim90$~h and 137~keV photons in only 9\% of the cases. To this aim, the design of the detector needs to take into account the existence of a  $\gamma$ radiation, albeit limited, and the softer $\beta^-$ spectrum (endpoint at 1.1~MeV as opposed to 2.3~MeV).

Finally, the requirements on the detecting probe are driven by the surgical conditions and by the characteristics of the radio-tracers: the probe needs to be sensitive to residuals as small as 0.1~ml, benchmark chosen by the sensitivity of the pre-operative diagnostics; it needs to be directional, i.e. insensitive to radiation from the sides; it needs to have the largest possible detection efficiency on electrons with the smallest sensitivity to photons; the smallest is the minimum electron energy that can be detected, the wider is the applicability of the technique because it allows for more radio-isotopes to be used and because it retains the sensitivity to tumor even if a thin layer of other tissues or liquids covers it; finally, it needs to match the security standards for intra-operative use.

In this context, this paper describes the design details of our first prototype and the test in laboratory of its performances.
Such tests had to be designed to reproduce by means of phantoms and automated procedures conditions as close as possible to the surgical environment. 
The outcome of these tests were input to the simulations that were used to estimate the sensitivity of the RGS technique proposed in several clinical cases~\cite{MeningiomaRGS,RENET}, studies that are preliminary to the clinical tests.

\section{The $\beta^-$ intraoperative probe}

In the first prototype of the intraoperative probe (Fig.~\ref{fig:probeS4})
the radiation sensitive element
is a scintillator tip 
made of commercial poly-crystalline para-terphenyl 
doped to 0.1\% in mass with diphenylbutadiene~\cite{Budakovsky}
manufactured by Detec-Europe.
This material was adopted, after a detailed study~\cite{PTerf},
due to its high light yield 
($\sim$3 times larger than typical organic scintillators),
negligible hygroscopy,
and low density, characteristics  that minimize the sensitivity to photons.
Different detector sizes were tested:
the best configuration resulted in a cylinder of 5~mm in diameter
and 3~mm in height.

The scintillator tip is shielded from radiation coming from the sides by
a black PVC ring with external diameter of 11~mm.
A 10~$\mu$m-thick aluminium front-end sheet
covers the detector window to ensure light tightness.
This assembly is mounted on top of
an easy-to-handle aluminum cylindrical body 
(diameter 8~mm and length 14~cm).
 
In order to avoid the risk 
of patients coming into contact with electrical devices,
the scintillation light is guided by four 50~cm long optical fibres
outside the probe to a Hamamatsu H10721-210
photo-multiplier tube (PMT).
This photo-sensor module has 
an integrated high voltage power supply circuit 
requiring an input voltage as low as 5~V, 
making this device compatible with the surgical environment.
Finally, the read-out electronics and the logic board are housed 
in a compact box that wirelessly connects to a remote monitor 
to display the counting rate~\cite{ardusipm}.

\begin{figure*}[bhpt]
 \centering
    \includegraphics[width=0.4\textwidth]{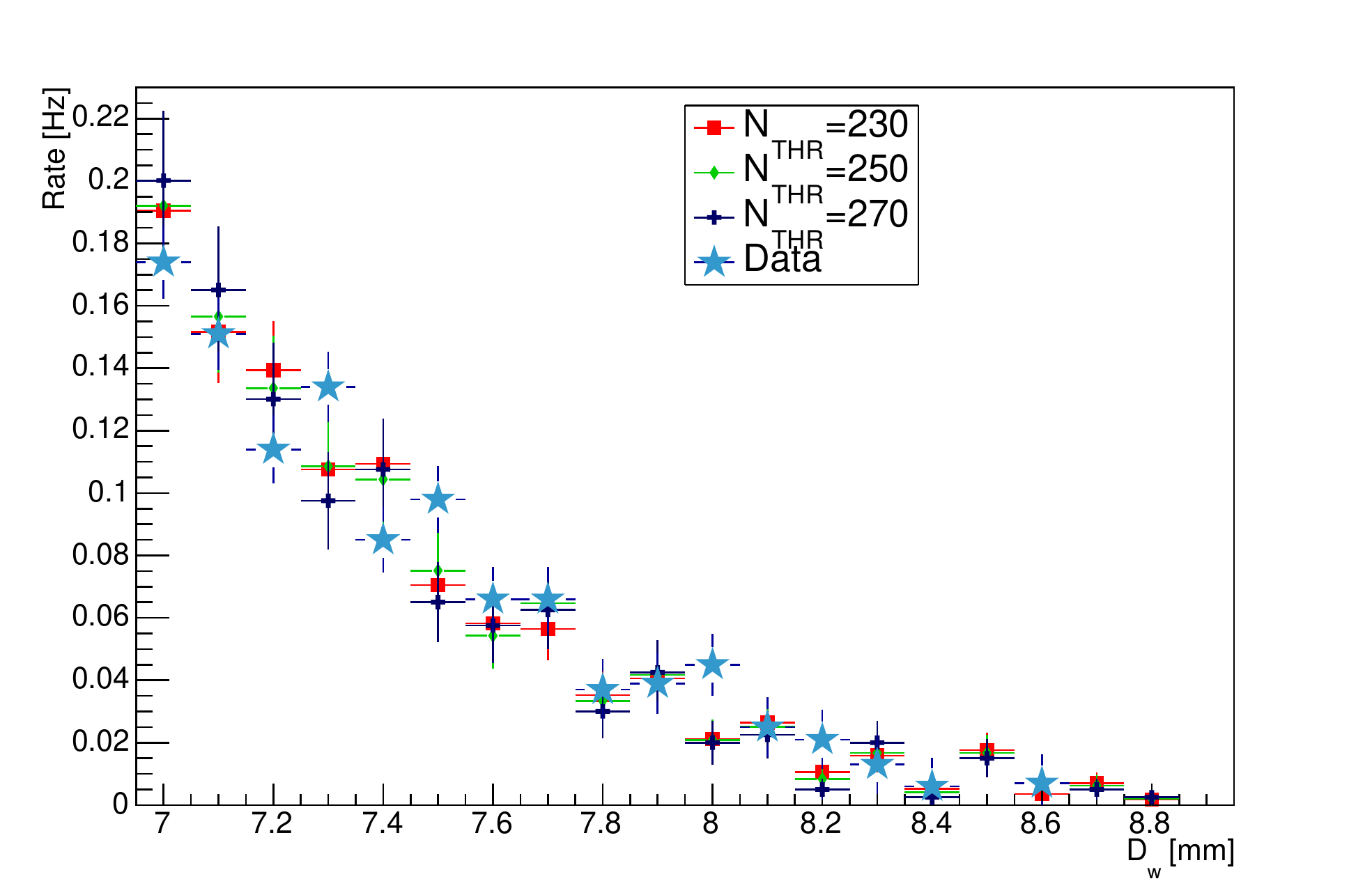}
    \includegraphics[width=0.4\textwidth]{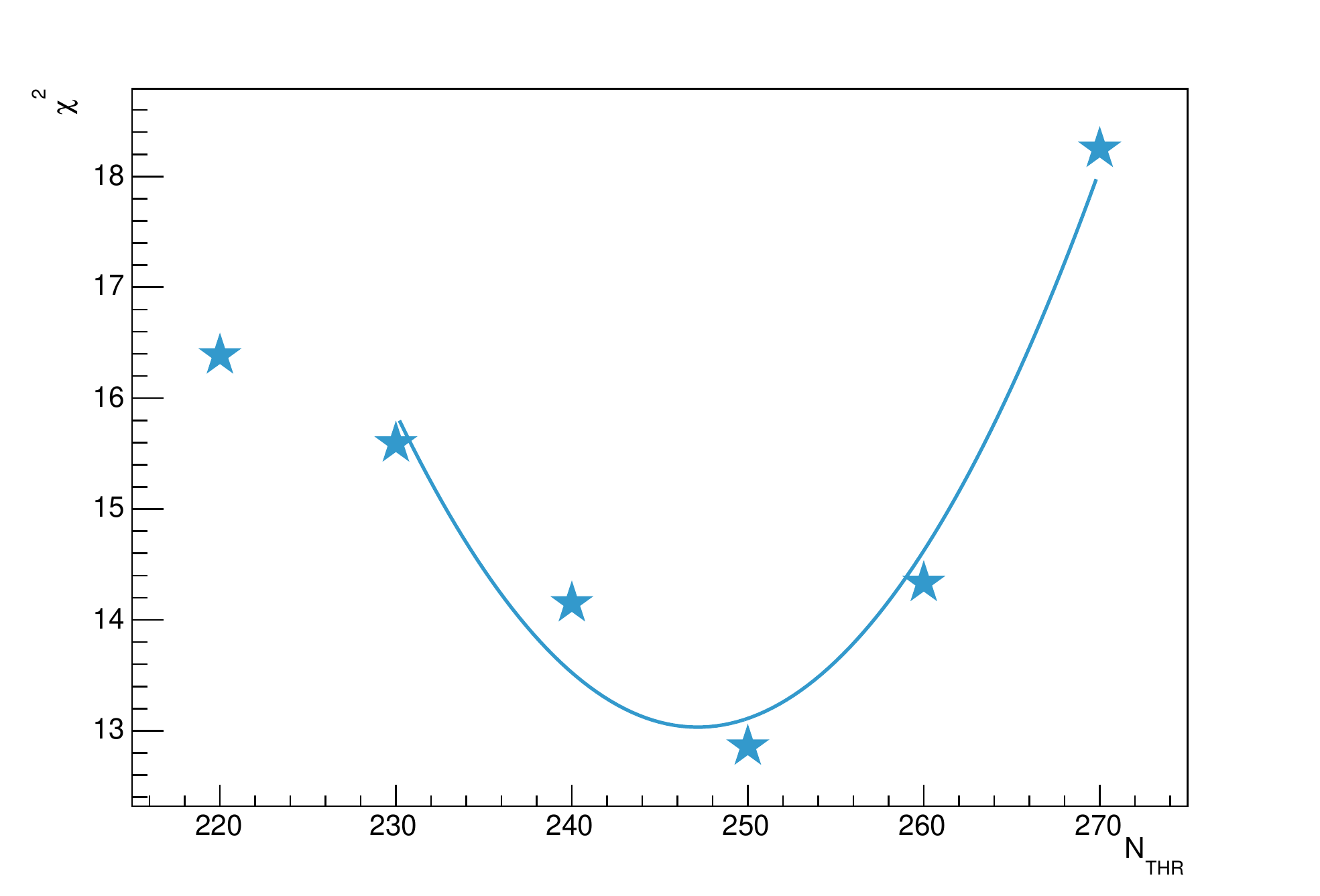}
    \caption{Left: Dependence of the electron rate as a function of water thickness $D_w$, in data and in several MC configurations differing by the threshold in the number of photoelectrons. Right: $\chi^2$ of agreement between data and MC (see text) as a function of $N_{thr}$ ($N_{thr}$). }
    \label{fig:WaterZoom}
 \end{figure*}

\section{Characterization of the $\beta^-$ intraoperative probe}
\label{sec:sealed}
The characterization of the probe response to $\beta^-$ decays
is performed with two  $^{90}$Sr sources
 a point-like one with nominal activity of $\sim$370~Bq
(\textit{point-source}) and a 1.6~cm diameter source with a $2.54\pm 0.15$~kBq activity (\textit{extended-source}). The $^{90}$Sr spectrum is made of two components in secular equilibrium: the Sr $\beta^-$ decay to $^{90}$Y and the subsequent  $\beta^-$ decay of the daughter. The latter component is exactly the one relevant for the proposed for RGS if $^{90}$Y is used.

\subsection{Efficiency}
To reduce the required injected activity it is important to maximize the efficiency. Furthermore, given the fact that even if we are currently considering $^{90}$Y as radionuclide different emitters will have to be considered in the future, it is important to know how low in energy the sensitivity of the probe extends. Therefore, we had to setup an ad-hoc measurement to evaluate the efficiency as a function of the emitted electron.

To this aim a vertical scan over the \textit{extended-source} was performed
measuring the rate with both the source and the probe in water
for several distances ($D_w$) between the two.
The presence of increasing amounts of water dumps 
the $\beta^-$ energy spectrum in a way that depends on the energy dependence of the efficiency. On the other hand, the absolute measured rate, and in particular the rate measured at contact (zero water thickness) depends on the absolute efficiency.
 %QUI
 \begin{figure}[tbhp]
 \centering
    \includegraphics[width=0.48\textwidth]{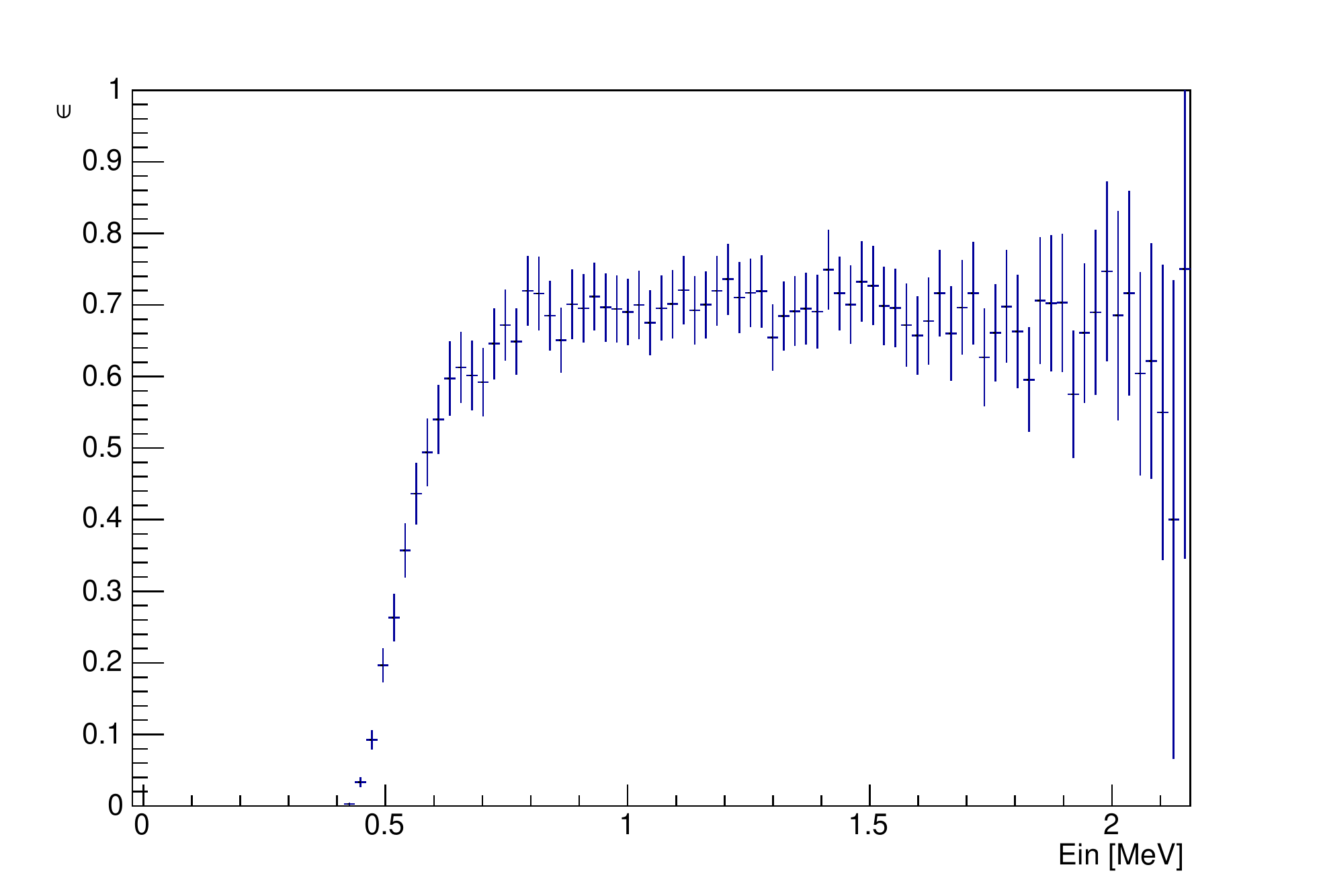}
    \caption{Probe efficiency as a function of the electron energy (E$_{in}$). }
    \label{fig:EffEne}
 \end{figure}

To correlate the measured rates with the efficiency 
a Monte Carlo (MC) simulation based on FLUKA code~\cite{FLUKA} was used. 
The FLUKA software fully simulates
the energy and position spectrum emitted by the  source,
the energy deposition in water and the probe. It also simulates the light emission and collection with the parameters measured in Ref.~\cite{PTerf}, returning the number of photoelectrons reaching the PMT ($N_{PMT}$). 
The fraction of  the electrons that are emitted by the source and reach the detector is estimated by the fraction of the simulated electrons that satisfy the condition  $N_{PMT}>N_{thr}$. It has to be stressed that this is not an estimate of the minimum number of photo electrons that are required to have a signal since the quantum-efficiency needs to be taken into account.

The comparison of this estimate with data  allowed to extract a value of $N_{thr}$, in two independent ways. On one side, $N_{thr}$ was tuned to match the measured rate in the configuration where the probe is in contact with the source. The best compatible threshold value was found to be $N_{thr}=240$.
On the other side, $N_{thr}$ affects the dependence of the rate with $D_w$ (see Fig.~\ref{fig:WaterZoom} on the left). A $\chi^2$ of agreement in between the data and MC rate variations between $D_w=7$ and $D_w=8.2$~mm was calculated for each value of $N_{thr}$.
Such $\chi^2$ showed a clear minimum in $N_{thr}=250$ (Fig.~\ref{fig:WaterZoom} on the right).

The two estimates of $N_{thr}$ are compatible and therefore we considered the average $N_{thr}=245$. To translate this estimate in terms of efficiency as a function of energy, we performed another simulation with the {\it point-source} and computed the ratio of the electron energy spectrum with and without the requirement $N_{PMT}>N_{thr}$.  The result is shown in Fig.~\ref{fig:EffEne}. The efficiency raises at energies above 540 keV (energy at which the curve reaches half of its maximum) and reaches the high energy plateaux at 70\%.

\subsection{Spatial resolution}

To study the capability of the probe
to reconstruct active spots, 
the detector was mounted on a motorized linear actuator
ensuring position accuracy of 1.5~$\mu$m and
horizontal scans were performed over the {\it point-source} in air.

 \begin{figure}[htbp]
\centering
 \includegraphics [width=0.48\textwidth]{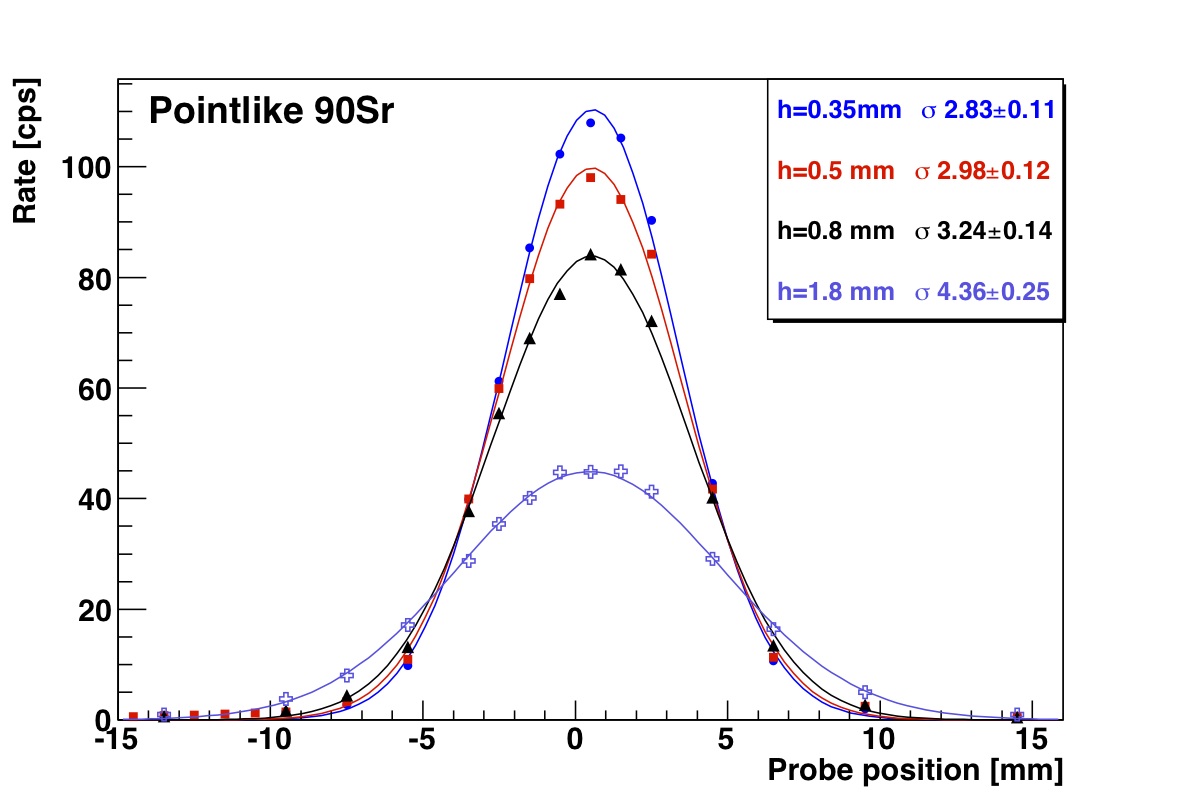}
 \caption{Profiles of the point-like \textit{Sr-source} 
 reconstructed by the probe at different distances
 from the source.}
 \label{fig:SrProfile}
 \end{figure}
An horizontal scan over the point-like \textit{Sr-source} 
(Fig.~\ref{fig:SrProfile}) allowed to estimate the dependence of the 
spatial resolution  on the distance 
between the probe tip and the surface.
The reconstructed profiles 
measured at distances ranging from 350~$\mu$m to 1.8~mm
are fitted using a Gaussian distribution
obtaining $\sigma$ between 2.8 and 4.4~mm.

%  \begin{figure}[htbp]
%\centering
%  \includegraphics [width=0.48\textwidth]{fig/ProfileHotSpotNoTitle.png}
%  \caption{Lateral scan of a 1.4~kBq, 0.1~ml, $^{90}$Y phantom. 
%  Gaussian and linear fits are
%  for the signal and dark count rate respectively.
%  See text for the band definition.}
%  \label{fig:HotSpotProfile}
% \end{figure}
%The scan over the extended source (Fig.~\ref{fig:HotSpotProfile}),
%instead, explores the discovery power of the probe
%in a more realistic simulation of a radio-labelled tumour residual.
%A cylindrical phantom with volume of 0.1~ml
%(diameter: 6.0~mm, height: 3.5~mm)
%is filled with a saline solution with $^{90}$Y radionuclide
%of 1.4~kBq activity,
%comparable to those administered for diagnostic purposes.
%The profile reconstructed by the probe is obtained
%with an horizontal scan of 1.5~mm steps and 10~s per position and
%a distance between the probe and the phantom surface of 100~$\mu$m.
%The probe is able to correctly identify the active spot 
%with a signal discriminating threshold S/N$>$5,
%where S is the number of detected events and
%N is the number of dark counts, and
%the real phantom dimension is reconstructed with S/N$>$20.
%

\subsection{Background rejection capability}
 
\begin{figure}
\centering
\includegraphics[width=0.48\textwidth]{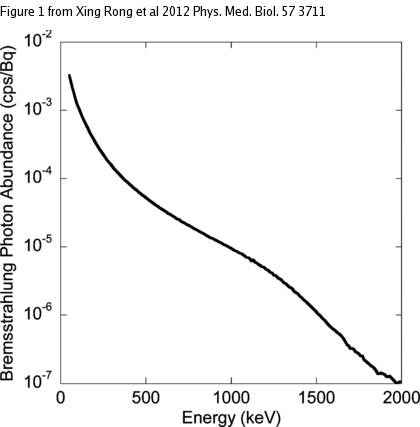}
\caption{Energy spectrum (50-2000~keV, 10~keV interval) 
of $^{90}$Y \textit{Bremsstrahlung} photons in water.}
\label{fig:Bremm90Y}
\end{figure}

The usage of pure $\beta^-$ emitting tracers
allows to operate with low radiation background, 
the residual being mainly due to photons 
coming from the \textit{Bremsstrahlung} radiation
of the electrons penetrating the tissue.
The abundance of the expected \textit{Bremsstrahlung} radiation 
in the tissue 
as a function of the photon energy 
is shown in Fig.~\ref{fig:Bremm90Y}
as computed with a simulation of
a $^{90}$Y source in water~\cite{90YBrem}. It is important in any case to ascertain that such radiation does not alter the locality of the measurement.
Conversely, in the case non pure $\beta^-$ emitters are used, it is important to minimize the sensitivity to $\gamma$s.

\begin{figure}[htbp]
\centering
 \includegraphics [width=0.48\textwidth]{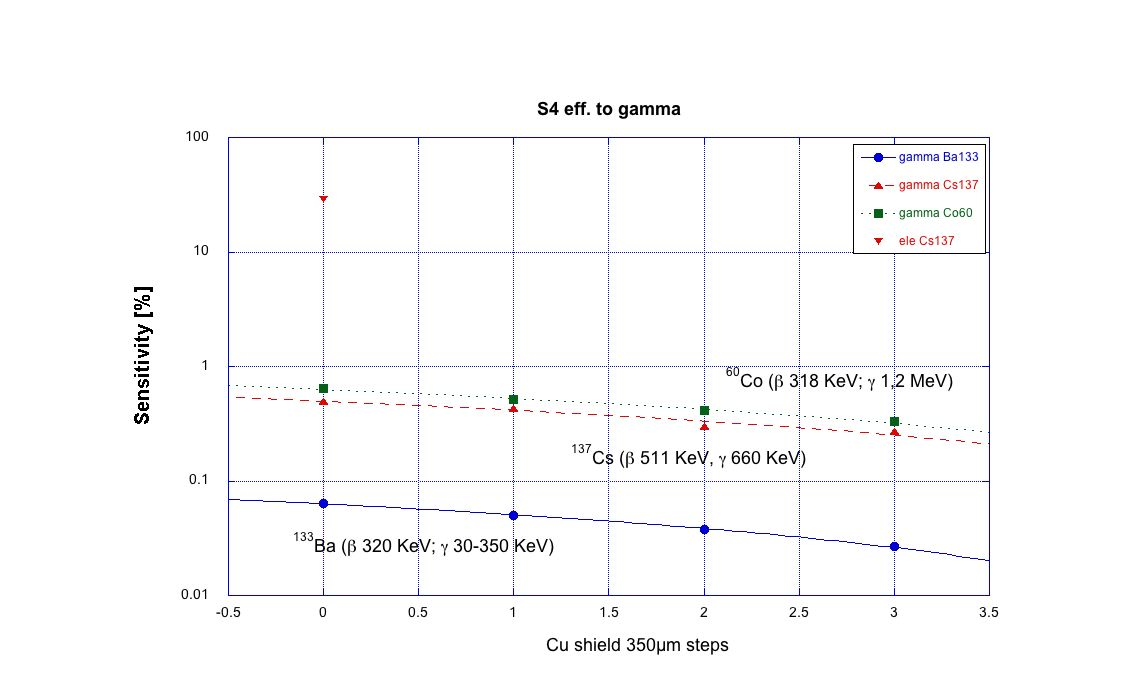}
\caption{Probe sensitivity to photons
emitted by three different sources:
$^{60}$Co E$_{\gamma1}$=1170~keV E$_{\gamma2}$=1330~keV, 
$^{137}$Cs E$_\gamma$=662~keV,
$^{133}$Ba E$_\gamma$ ranging from 80 to 350~keV.
The measurements were repeated after insertions of
copper layers with thickness of 350~$\mu$m
between the source and the probe tip 
to absorb the electronic component
of the Caesium emissions.}
\label{fig:Bremm90Yeff}
\end{figure}

To this aim, the sensitivity of the probe to photons 
was measured using three point-like sources:
$^{133}$Ba emitting photons with energy ranging from 80 to 350~keV,
$^{137}$Cs with gamma emission at E$_\gamma$=662~keV and
$^{60}$Co  with E$_{\gamma1}$=1170~keV and E$_{\gamma2}$=1330~keV.
To avoid signal from electrons in the case of Cs decays,
three copper layers with 350~$\mu$m thickness were inserted in sequence
between the source and the probe tip,
and the measurements were repeated at each step.
The counts as measured by the probe are shown 
in the Fig.~\ref{fig:Bremm90Yeff} for the three sources.
Except for the first measure on the Cs source,
introducing the copper absorbers implies a very small decrease in rate 
compatible with the attenuation in copper and
the change in geometrical acceptance.

The sensitivity to the photons emitted by the $^{133}$Ba source
is 0.06\%, 
whereas for the $\gamma$s from the $^{137}$Cs and $^{60}$Co sources
the sensitivity is %lightly higher but 
still 0.5\%.
These measurements allow us to conclude that 
the intraoperative $\beta^-$ probe is not sensitive to 
the \textit{Bremsstrahlung} photons
and therefore the effectiveness of the RGS technique would
not be affected by this background.

\section{Preclinical tests on specific phantoms}
\label{sec:phantoms}
A critical aspect in the test of the prototypes for the development of the new RGS technique, is the difficulty to perform realistic studies before the preclinical test. This aspect is particularly important for this technique because the outcome of the laboratory test is used to evaluate which clinical cases to consider.
To this aim, specific phantoms were designed to
reproduce finite size meningioma remnants 
embedded in healthy brain tissue
with a TNR as expected in a real clinical case, namely meningioma that is considered the test bench of this technique. 
The same set-up allows going through different kinds of interfaces and
visual/acoustic feedback to identify which one could best assist 
the surgeon in the search for infected residuals.
 
\subsection{Realization of the phantoms}
 
The surgical environment is simulated saturating conveniently shaped 
cuttings of a commercially available sponge 
(Wettex  Classic by Vileda$^{\textregistered}$) 
with  $^{90}$Y-DOTATOC in saline solution.

The sponge is made of 65$\%$ cellulose and 35$\%$ cotton fibres 
packed as 20$\times$20~cm wide and 2~mm thick dry sheets. 
The choice of the sponge was driven by its high 
water absorbency (as measured by us on small samples),
the ease of obtaining a sharp and precise cut even with 
few millimetre wide complex shapes, and the capability of a sample to regain its original dimensions  
after a wetting-drying cycle.

%To radio-protect operators,
%the phantoms are set up and used within a 10~mm thick PMMA glove box. 

With this technique, both wet or dry surfaces can be produced 
depending on the scope of the test: dry phantoms are used to test the neat physics detector performances
(e.g. sensitivity to active spots, capability of profile reconstruction); on the contrary, 
wet surfaces are suitable 
for studying and optimizing the probe usage 
in a more realistic environment where, for example, 
tip contamination might occur adding a random background 
to the signal coming from the tissues.

\begin{figure}[htbp]
\centering
\includegraphics [width=0.3\textwidth]{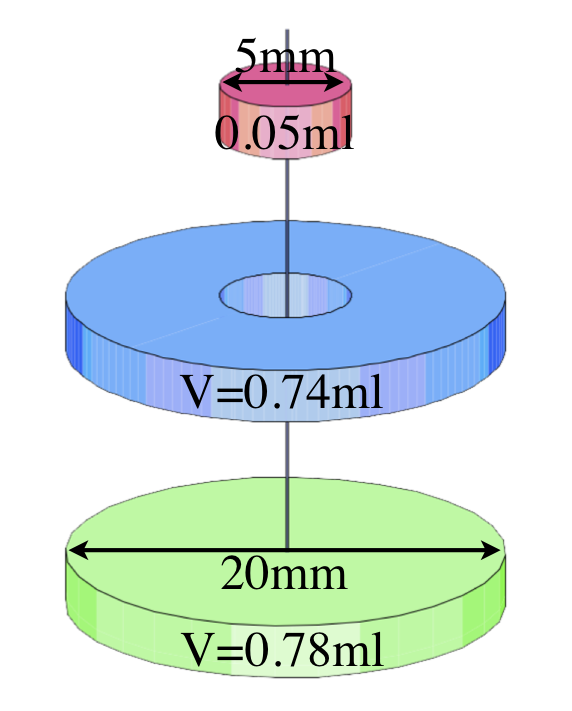}
 \caption{Assembling of a phantom simulating 
 a \textbf{tumour residual} of 0.05~ml 
 embedded in surrounding (the \textbf{ring}) 
 and underneath (the \textbf{disk}) healthy tissue.}
 \label{fig:BestaRingPhantom}
\vspace{0.5cm}
\includegraphics[width=0.48\textwidth]{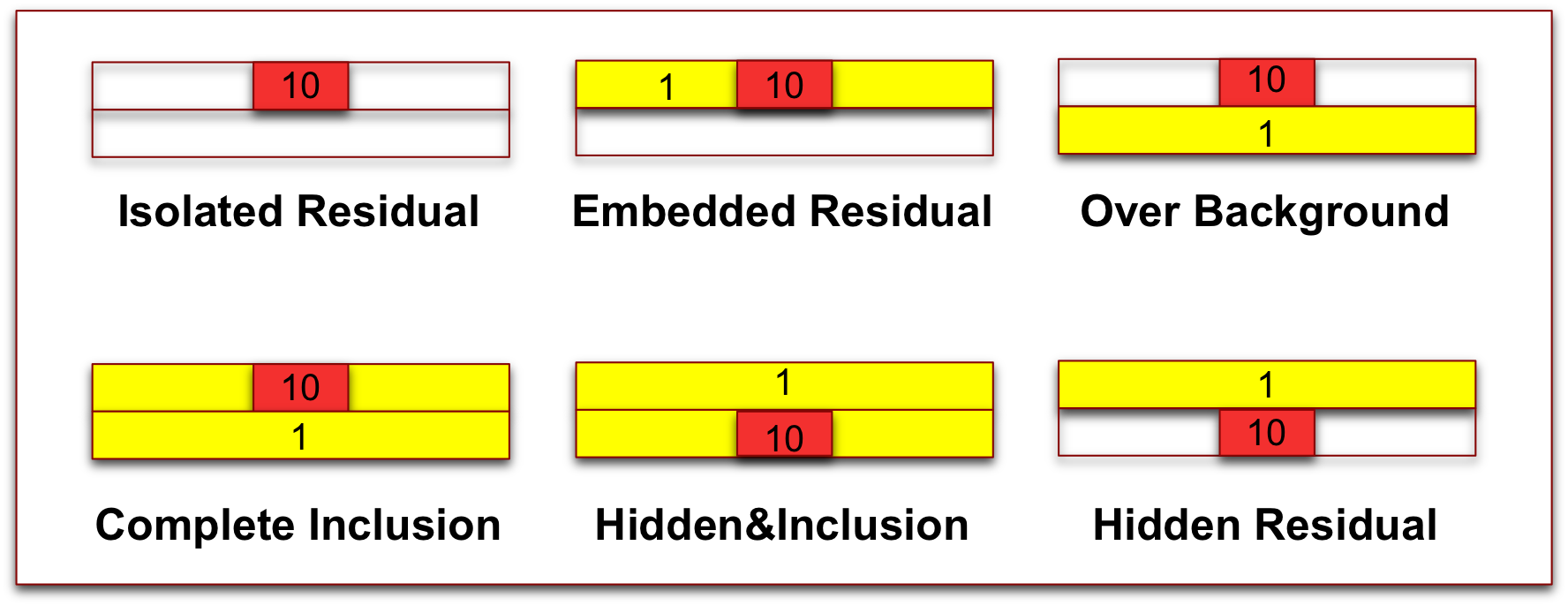}
 \caption{Varying position and activity 
 (white: non-active tissue, yellow: active tissue) 
 of the assembly elements 
 the discovery potential of the probe can be tested on surfaces 
 reproducing, even if schematically, some possible combinations 
 of cancerous residual location and TNR occurring during surgery.}
 \label{fig:BestaRingComb}
 \end{figure}

The simplest assembly made with this technique is shown 
in Fig.~\ref{fig:BestaRingPhantom}:
a 5~mm diameter circular cutting 
simulating a 0.05~ml tumour residual
is inserted inside a larger (20~mm diameter) ring 
reproducing the nearby healthy tissue.
A third round shape (disk) was eventually used to reproduce 
the presence of further healthy tissues above or under the hot spot. 
Different TNR between the tumour and the surrounding healthy tissue 
were realized drenching the cuttings, up to saturation, 
with the properly diluted $^{90}$Y solution.  
The elements were then dried and combined 
to shape the topology of interest (Fig.~\ref{fig:BestaRingComb}).
 
The nominal activity of any phantom was inferred 
from the activity of the $^{90}$Y-DOTATOC vial 
at the time of delivery and 
the dilution factor of the bath prepared to drench the samples. 
The relative activity of the final dried phantoms was individually 
measured before forming the assembly.

\subsection{Active spot identification}

A scan over the described assembly was performed by fixing the probe 
to a XZ motorized system. 
The X linear actuator run the probe over the phantom in 1~mm steps, 
with a measurement time of 10~s per position,
while the Z actuator was used to set the tip to surface distance 
with a precision of 1.5~$\mu$m.
 
The different configurations were analyzed 
in terms of false-positive probability (FP), 
i.e. the probability that the rate emitted 
by an healthy tissue is incorrectly flagged by the probe 
as tumour residual.
In the actual case, 
this flag will be set in real time by the system 
comparing the signal measured over the tissue under investigation and 
the patient reference background 
previously measured over their healthy tissue. 
The probability of false-positive then depends on the probe efficiency, 
the time to response and the actual TNR.

\begin{figure}[htbp]
\centering
 \includegraphics [width=0.48\textwidth]{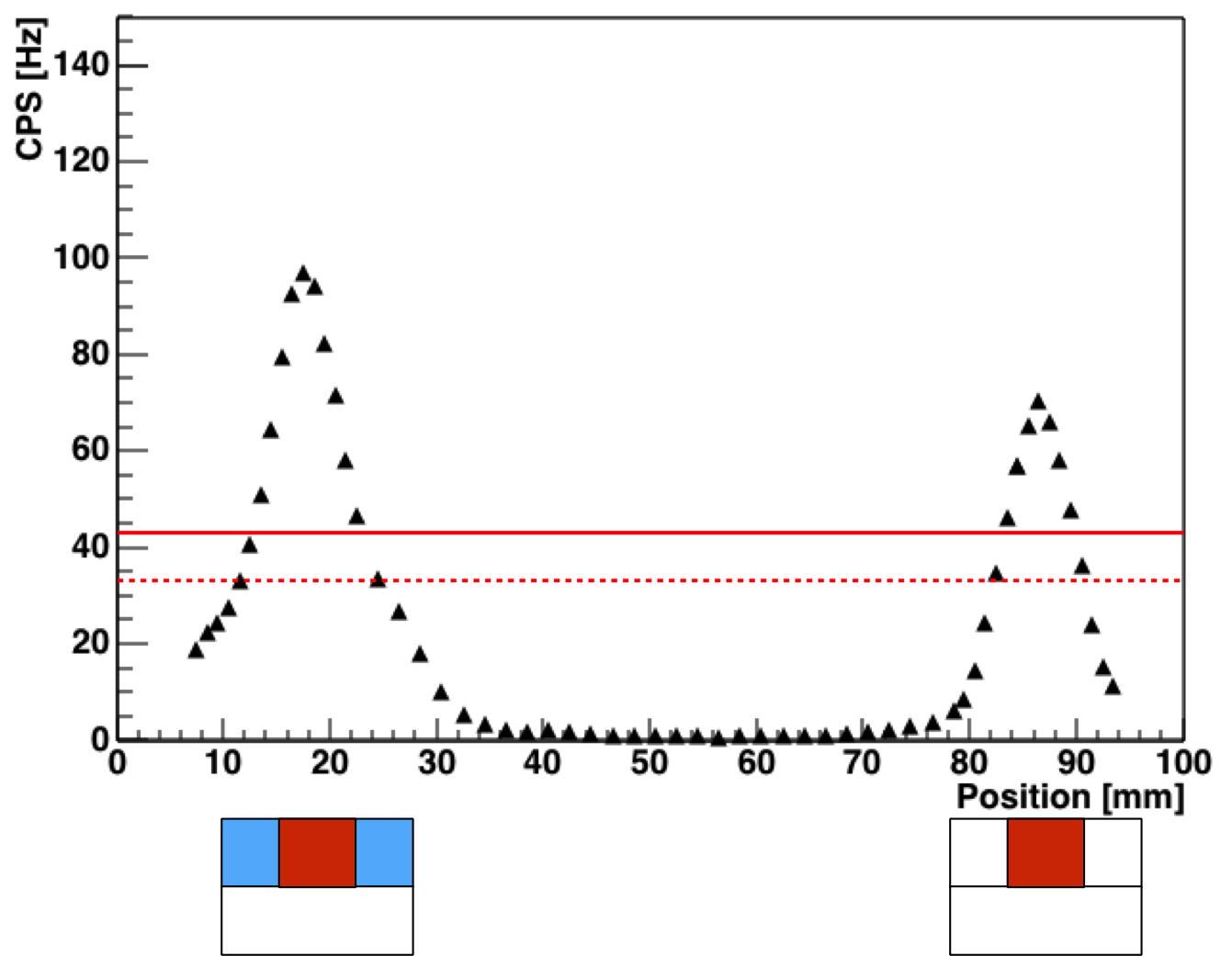}
  \caption{Scan over two assemblies with a probe tip-surface distance 
  next to zero. 
  The left phantom simulates a 0.5~ml tumour surrounded 
  by a same thickness healthy tissue with a TNR of 10
  (as observed in patients affected by meningioma), 
  the second assembly is an isolated residual 
  (non-active surrounding tissue).The full (dashed) line corresponds to the rate needed to ensure a 1\% false positive rate in 1 (10) seconds. See the text for details on the source representation below the plot.}
 \label{fig:scan1}
 \end{figure}

An example of a scan  is shown in Fig.~\ref{fig:scan1}, where
the horizontal lines indicate the minimum rate needed to ensure FP$<1\%$ with a measurement. Below the figure a scheme of the source indicates if the individual regions have the activity associated to background (blue area) or 10 times more (red area). The unfilled area corresponds to non-active regions.
The device clearly selects the correct region. 
 
\begin{figure}[htbp]
 \centering
 \includegraphics [width=0.48\textwidth]{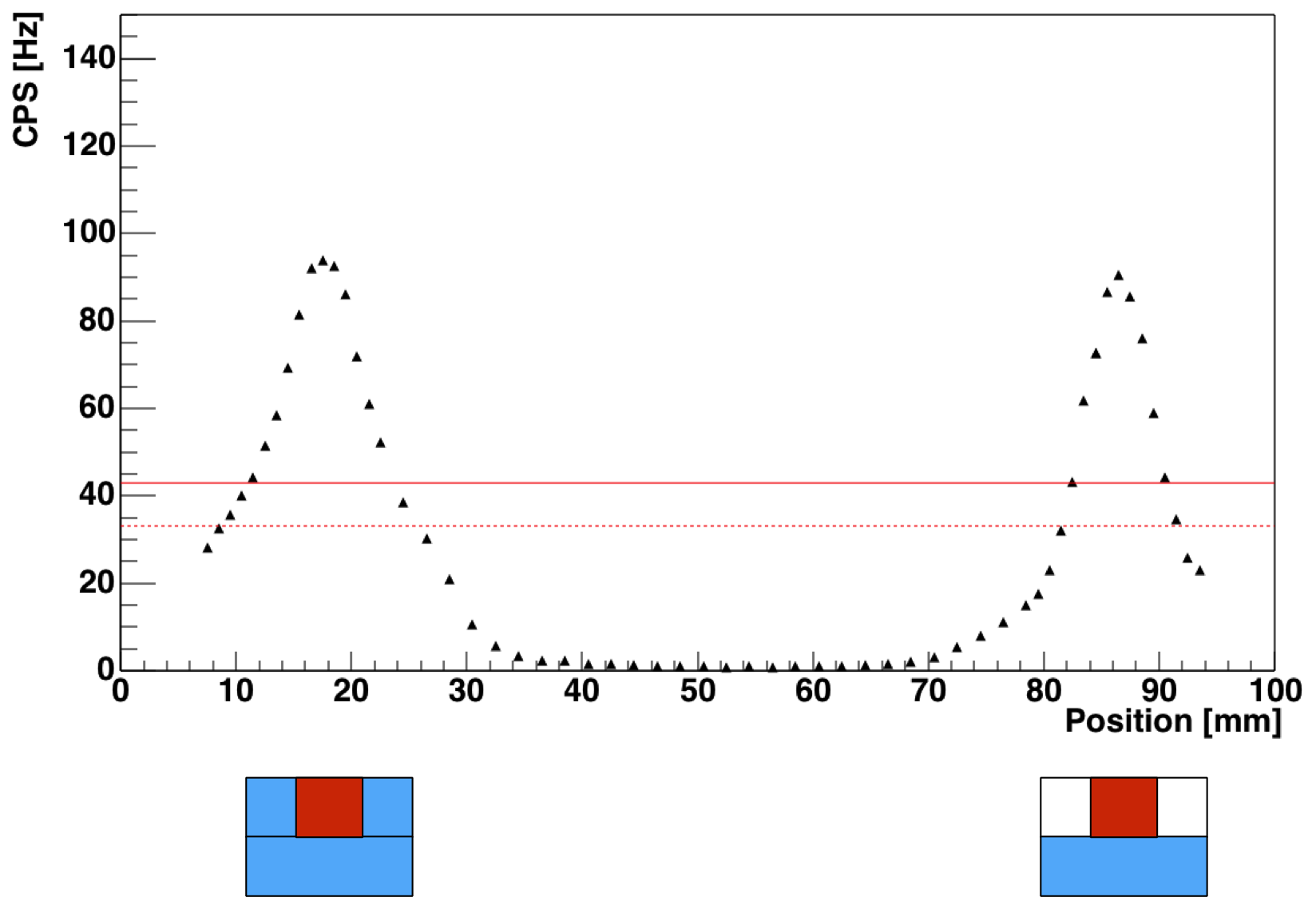}
  \caption{Scan over two assemblies 
  with a probe tip-surface distance close to zero 
  realized to study the sensitivity of the probe to a 0.5~ml hot spot 
  masked by side and beneath emitting (TNR=10) healthy tissue. The full (dashed) line corresponds to the rate needed to ensure a 1\% false positive rate in 1 (10) seconds. See the text for details on the source representation below the plot.}
 \label{fig:scan2}
 \vspace{0.5cm}
%\end{figure}
% 
% \begin{figure}[htbp]
% \centering
 \includegraphics [width=0.48\textwidth]{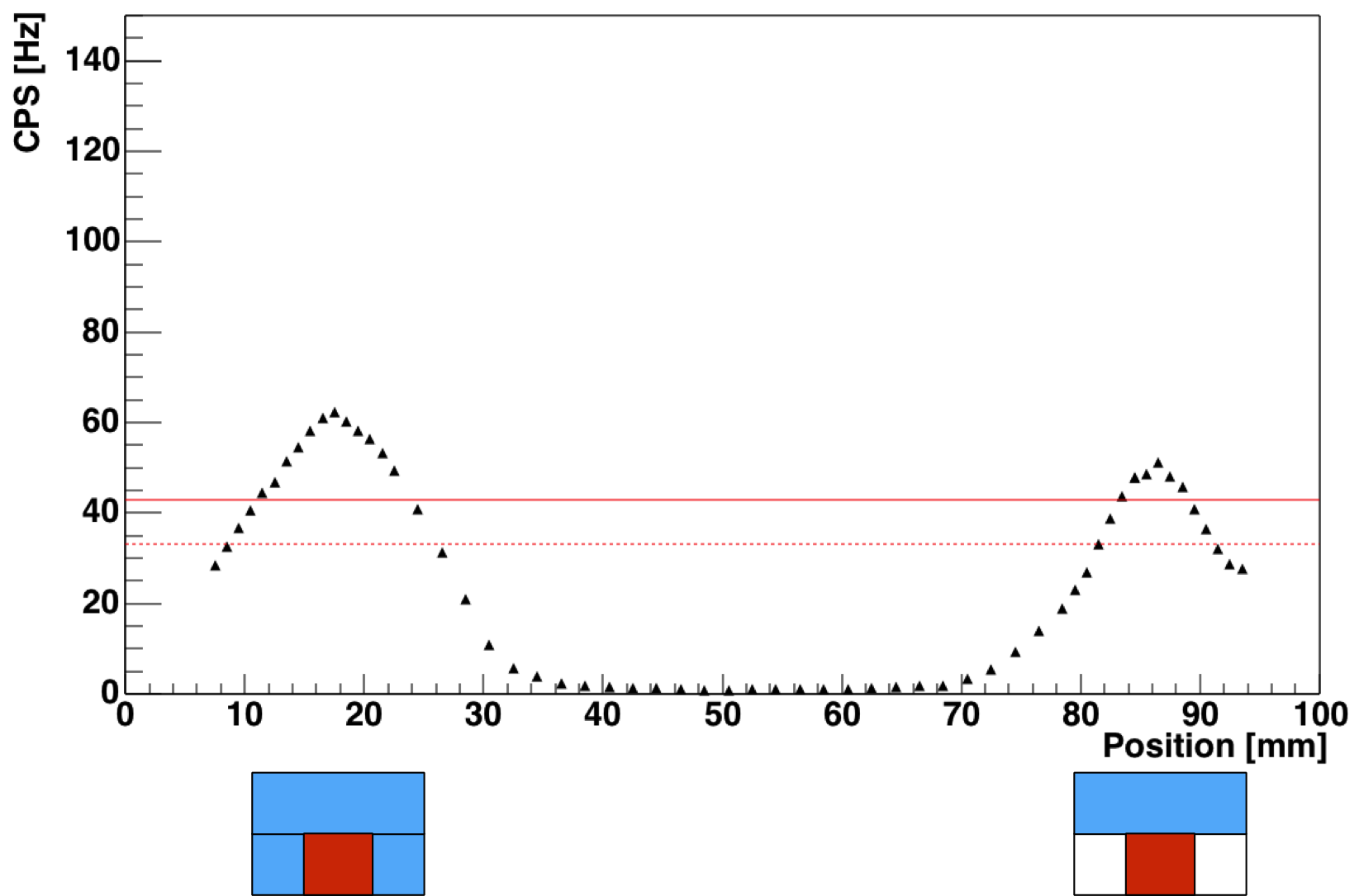}
  \caption{Scan over two assemblies with a probe tip-surface distance
  equal to zero to test the sensitivity of the probe to a hot spot 
  hidden under an emitting (TNR=10) healthy tissue. The full (dashed) line corresponds to the rate needed to ensure a 1\% false positive rate in 1 (10) seconds. See the text for details on the source representation below the plot. }
 \label{fig:scan3}
 \end{figure}

Fig.~\ref{fig:scan2} and~\ref{fig:scan3} show the same measurement
performed on different tumour-health tissue patterns 
(fixed the specific activity and the TNR) 
that might affect the discovery potential of the probe. 
 
These evaluations were performed 
in a neat and well known configuration,
whereas the surgical environment might introduce some perturbations 
as, for example, a contamination of the probe tip 
coming into contact with the wet active tissue. % and
Such effect can be evaluated directly from the scan 
shown in Fig.~\ref{fig:scan1} 
adding, via software, an extra background to the measure up to 
the minimum TNR that makes still detectable the hot spot. 
The study lead to the conclusion that the probe is able to correctly identify
the active spot also in case of TNR as low as 2
with $FP=1\%$ C.L. and a time of measure of 10~s.
That means that an eventual contamination 
would not be a problem if monitored
with frequent background calibration during the operation.
 
Moreover, 
these results imply that the RGS technique with $\beta^-$ decays
would be effective also in case of patients with a lower TNR
compared to those affected by meningioma,
and therefore they are a confirmation that the RGS could be extended
to clinical case of interest with an eventual uptake 
of the radiotracer in the nearby healthy organs~\cite{MeningiomaRGS}.

\subsection{Minimum detectable tumour residual}
 
\begin{figure}[htbp]
\centering
 \includegraphics [width=0.4\textwidth]{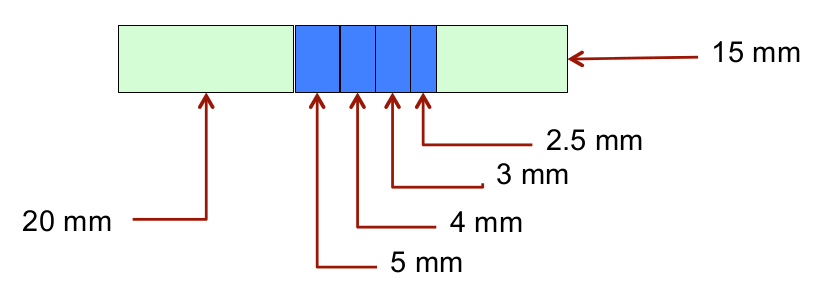}
  \caption{The phantom prepared to simulate the progress 
  of a surgical removal of a tumour is made by a strip (blue)
  of 4 tiles of different sizes and charged at the tumour uptake.
  A sequence of scans was performed 
  removing a strip at each iteration
  to study the minimum detectable residual.
  At the beginning and the end of this assembly two tiles (green)
  ten times less active define the boundary of the healthy tissue 
  surrounding the tumour.}
 \label{fig:strip1}
  \end{figure}

A different phantom pattern was used to simulate and 
study the evolution of the probe response 
while removing an extended tumour area aiming at the identification 
of the minimum detectable residual.
For this purpose the assembly shown in Fig.~\ref{fig:strip1} was used.
A scan over the whole strip was repeated reducing the tumoural area
by removing a tile at a time starting from the wider one (5~mm). 
At each iteration the assembly was packed again to always 
keep the tumour area enclosed between healthy tissues.
 
\begin{figure}[htpb]
 \centering
 \includegraphics [width=0.48\textwidth]{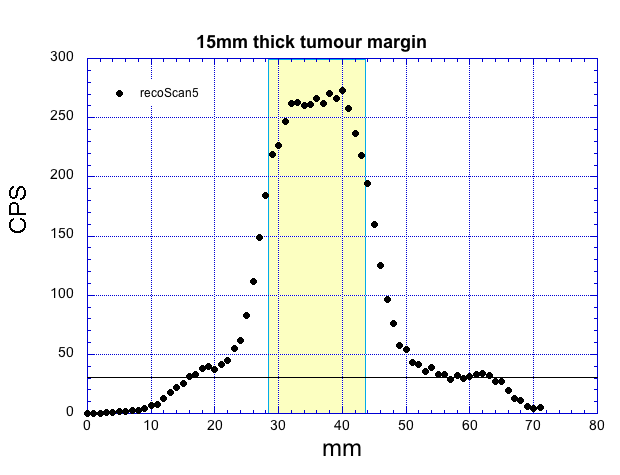}
  \caption{A scan over a 15$\times$10~mm$^2$ tumour activated phantom. 
  The yellow area indicate the position of the tiles 
  between the two healthy tissue phantoms. 
  The limited lateral shield makes the probe sensitive to the approaching 
  active area about 4~mm before the detector goes over the hot tiles.}
 \label{fig:15mm_tumour_margin}
  \vspace{0.5cm}
% \end{figure}
% \begin{figure}[htpb]
%  \centering
  \includegraphics [width=0.48\textwidth]{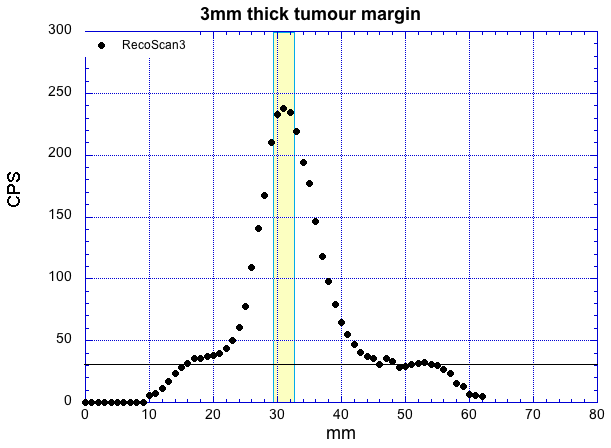}
   \caption{Scan over a 3~mm wide tumour tile phantom. 
   Since the width of the phantom is smaller than the
   sensitive detector dimension,
   the distribution shows a peak 
   corresponding to the alignment centred over the sample.}
  \label{fig:3mm_tumour_margin}
   \vspace{0.5cm}
%  \end{figure}  
% \begin{figure}[htpb]
%  \centering
  \includegraphics [width=0.48\textwidth]{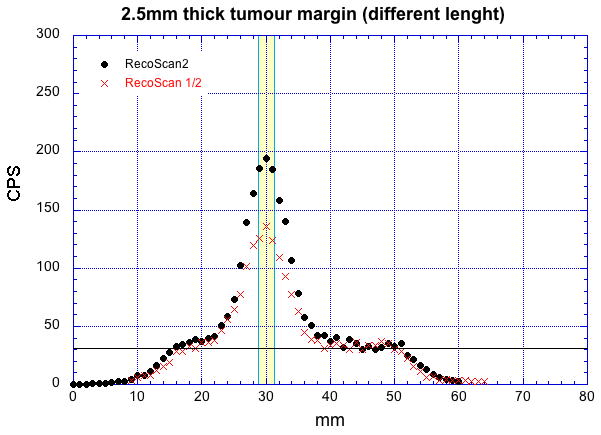}
   \caption{Scan over a 2.5$\times$10~mm$^2$ tumour margin. 
   The red cross refers to the same scan performed on a 
   smaller 2.5$\times$5~mm$^2$ sample.}
  \label{fig:2_half_tumour_margin}
  \end{figure}

The scan over the full 15~mm long tumour phantom is shown in 
Fig.~\ref{fig:15mm_tumour_margin}. 
Following the scan from the right to the left 
the experimental shape is read as follow:  
the cps counts start from zero (the probe is outside the scanning area) 
and increase as the probe goes over the healthy tissue. 
The small plateau at 30~cps defines the background level of this measure. 
Approaching the tumour phantom, 
the counts increase up to plateau at about 250~cps 
lasting as much as the probe sensitive area is fully illuminated 
by the tumour area. 
As soon as the sensitive detector exits the last tumour tile 
the rate goes back to that of the healthy tissue.
  
The responses measured reducing the tumour thickness to 3~mm and 2.5~mm
are shown in Fig.~\ref{fig:3mm_tumour_margin} and 
Fig.~\ref{fig:2_half_tumour_margin} respectively. 
In both cases the tumour margins are clearly identified 
once the activity of the surrounding healthy tissue is known. 

The smallest residual tested in this measure, 
obtained reducing by 5~mm the length of the 2.5~mm tile, 
has a volume of 0.03~ml.
The corresponding signal is shown in
Fig.~\ref{fig:2_half_tumour_margin}.

%\subsection{Human feed-back}

\section{Conclusion}

We reported on the performances of the first prototype 
of the intraoperative probe 
purposely designed to exploit the newly proposed
radio-guided surgery with $\beta^-$ decays.
The low density and the high light yield of 
the para-terphenyl scintillator 
allow the development of a very compact device
with a good sensitivity to electrons in the range of the $^{90}$Y decays 
and almost transparent to the \textit{bremsstrahlung} photons.
The first results obtained experimenting the probe 
on point-like and extended $^{90}$Y sources 
highlight the potentiality of this approach
in all the issues relevant in this field: 
millimetric remnants discovery potential, spatial resolution, 
fast response, dose administered to the patient, 
operators safety.

% % %\section*{Acknowledgment}

% that's all folks
\end{document}